\documentstyle[12pt]{article}
\newlength{\extraspace}
\newlength{\extraspaces}
\begin{document}
\addtolength{\baselineskip}{.7mm}

\thispagestyle{empty}

\begin{flushright}
{\sc KAIST CHEP} 95/17\\
%{\sc PUPT-xxxx}\\
%hepth@xxx\\
October 1995
\end{flushright}
\vspace{.3cm}

\begin{center}
{\Large\bf{General Static Solutions of 2-dimensional  
 \\[2mm]   Einstein-Dilaton-Maxwell-Scalar Theories 
}}\\[10mm]
{\sc Dahl Park}\\[3mm]
{\it Department of Physics\\[2mm]
 KAIST\\[2mm]
Taejon 305-701, KOREA\\[2mm] }
{\sc And}\\[3mm] 
{\sc Youngjai Kiem}\\[3mm]
{\it Department of Physics\\[2mm]
 Sejong University\\[2mm]
 Seoul 133-747, KOREA\\[2mm]
E-mail: dpark@chep6.kaist.ac.kr \\
        ykiem@phy.sejong.ac.kr}
\\[20mm]

{\sc Abstract}
\end{center}

General static solutions of effectively 2-dimensional
Einstein-Dilaton-Maxwell-Scalar theories are obtained.  Our
model action includes a class of 2-d dilaton gravity theories coupled 
with a $U(1)$ gauge field and a massless scalar field.  Therefore
it also describes the spherically symmetric reduction of 
$d$-dimensional Einstein-Scalar-Maxwell theories. The properties of the
analytic solutions are briefly discussed.
  
\noindent

\vfill

\newpage

\section{Introduction}

The model action we consider in this paper is given as follows. 
\begin{equation}
I = \int d^2 x \sqrt{-g} e^{-2 \phi} [ R + \gamma g^{\alpha \beta}
\partial_{\alpha} \phi \partial_{\beta} \phi + \mu e^{2\lambda\phi} 
-\frac{1}{2}e^{- 2 \phi (\delta - 1) } g^{\alpha \beta} \partial_{\alpha} f
\partial_{\beta} f + \frac{1}{4} e^{\epsilon \phi } F^2 ],
\label{oaction}
\end{equation}
where $R$ denotes the 2-d scalar curvature and 
$F$, the curvature 2-form for an
Abelian gauge field. $\phi$ and $f$ represent a dilaton field and 
a massless
scalar field, respectively. The parameters $\gamma$, $\mu$, $\lambda$,
$\epsilon$ and $\delta$ are assumed to be arbitrary real numbers. 
A specific choice of these 5 parameters corresponds to a particular
gravity theory.  Action (\ref{oaction}) is of interest in itself
as a 2-d dilaton gravity theory coupled with a $U(1)$ gauge
field and a scalar field, since it contains various couplings of
the dilaton field to other fields \cite{teitel}\cite{BL}.  In 
particular, the choice 
of $\gamma = 4$ and $\lambda = 0$ reduces the gravity sector of the action
to the theory of Callan, Giddings, Harvey and Strominger (CGHS) \cite{cghs}.
This string-inspired model has provided us with an analytically tractable 
framework to study the gravitational physics \cite{BL}\cite{giddings}. 
Additionally, our model represents the
spherically symmetric reduction of large class of $d$-dimensional
Einstein-Maxwell-Scalar theories \cite{birkhoff}.  From this point of view,
2-d dilaton field $\phi$ is directly related to the geometric
radius of each $(d-2)$-dimensional sphere in $d$-dimensional spherically
symmetric geometry. To be specific, we can write the
spherically symmetric $d$-dimensional metric $g^{(d)}_{\alpha \beta}$
as the sum of          
longitudinal part and transversal angular part 
\[ ds^2 = g_{\alpha \beta} dx^{\alpha}dx^{\beta} - 
\exp { ( - \frac{4}{d-2} \phi)} d \Omega^2 \]
where $d\Omega^2$ is the metric of a sphere $S^{d-2}$ with 
unit radius and we use $(+- \cdots -)$ metric 
signature.  The spherically symmetric reduction of      
$d$-dimensional Einstein-Maxwell-Scalar action 
\begin{equation}
 I = \int d^d x \sqrt{g^{(d)}} ( R - \frac{1}{2}
g^{(d) \alpha \beta} \partial_{\alpha} f \partial_{\beta} f
+ \frac{1}{4} F^2 )
\label{daction}
\end{equation} 
becomes Eq.(\ref{oaction}) with $\gamma = 4 (d-3)/(d-2)$,
$\lambda = 2/(d-2)$, $\delta = 1$, $\epsilon = 0$ and $\mu$,
a constant depending on the area of $(d-2)$-dimensional sphere 
after $(d-2)$-dimensional angular integration. If $d=4$, for example,
we have $\mu=-2$.  The spherically symmetric reduction of 
4-dimensioanl Einstein-Maxwell-Fermion theories, after
the bosonization of the Callan-Rubakov modes of the fermions,
also becomes (\ref{oaction}) with the above values of 
the parameters, except $\delta = 0$ in this case \cite{trivedi}.

     Ongoing debates on the quantum evaporation of black 
holes \cite{giddings} provide a major motivation for studying
the gravity theories described by the action (\ref{oaction}).
In this regard, it has been suggested that the possible end 
points of the black hole evaporation process are the quantum
deformations of the extremal solutions similar to the extremal 
Reissner-Nordstr\"{o}m space-time \cite{banks}.  A class of 
2-dimensional dilaton theories of the type (\ref{oaction}), 
where we have a $U(1)$ gauge field, are useful
frameworks for investigating this 
possibility \cite{banks}\cite{lowe}\cite{st}.  A key issue along
this line of investigation
is the inclusion of the (quantum) gravitational back reactions.

     An important step toward the understanding of those
gravity theories, therefore,
is to obtain their general classical solutions with exact treatment
of the classical gravitational back reactions caused by the $U(1)$
gauge field and the scalar matter field. 
Even when we are interested
in static solutions, this task is largely
hampered by the non-linearity of the classical equations of motion.
What we then need is a systematic procedure to solve this set of non-linear
coupled equations.\footnote{See \cite{rakh}, \cite{gs} and 
references cited therein for 
other approaches to this problem}  For some simpler
theories of
interest, the situation is more favorable.  In case of 4-dimensional 
Einstein-Scalar theory, the methods of Buchdahl \cite{buchdahl}
and Janis et.al. \cite{janis} are reported in literature.  More 
recently, Myers and Perry presented an extensive
study of $d$-dimensional Einstein-Maxwell theory \cite{myers}.

     In Ref.\cite{kp}, in the absence of $U(1)$ gauge field,
we utilized the static remnants of 
underlying classical conformal invariance to reduce the set of second-order
coupled ordinary differential equations (ODEs) to the set of first-order
ODEs.  The ODEs were summarized as the conservation of the corresponding 
Noether charges.  In the present work, we generalize that approach by 
including a $U(1)$ gauge field.  From this point of view, the main novelty
of our work is its method of derivation.  In what follows, we construct
4 Noether charges so that the 4 fields, namely, the conformal factor
of the metric
in conformal gauge, the dilaton field, the scalar field and the $U(1)$
gauge field, can be solved from a set of first-order ODEs.  This 
construction is possible for a general choice of 5 parameter values
introduced above.  Moreover, under the restriction of 
$2 - \lambda - \gamma /2 + \epsilon /2 = 0$ and $\delta = 1$, we can
go further and get the general static solutions in a closed form.
It is regrettable that this restriction is necessary for some technical
reasons when we try to obtain the solutions in a closed form.  However,
the spherically symmetric reduction of $d$-dimensional 
Einstein-Maxwell-Scalar theory and the CGHS model, two most important 
cases in our consideration, satisfy the restriction.  Finally, we 
discuss the properties
of our solutions and some aspects of classical back reactions
implied by our solutions.

     Recently, G\"{u}rses and Sermutlu derived general static spherically
symmetric solutions to $d$-dimensional Einstein-Dilaton-Maxwell theory
(with additional dilaton coupling to $U(1)$ gauge field that we do not
consider in this paper) \cite{gs}.
By integrating out the angular dependence from the outset to derive 
effectively 2-dimensional action, we circumvent some of the technical
complexity of their approach while getting the general static solutions
for the theories continuously connecting the 4-d $s$-wave Einstein theory
to the CGHS model.  Our approach also provides a method that works under
the choice of more conventional conformal gauge.  Still, it remains to be 
seen whether the analysis in our fashion can be used in deriving 
general static solutions in the presence of the coupling between
the dilaton field and $U(1)$ gauge field in $d$-dimensional theories
($d>2$), as dictated from the low
energy limit of string theory \cite{horowitz}.

\section{The Derivation of General Static Solutions}

     We start by deriving the static equations of motion.  Then the 
existence of symmetries is pointed out to explicitly construct the 
corresponding Noether charges.  Under the aforementioned restriction,
we can further integrate the ODEs to get closed form solutions.

\subsection{Static Equations of Motion}

     The equations of motion 
are obtained from the action by varying it with respect to the metric tensor,
the dilaton field, the massless scalar field and the gauge fields;
\begin{equation}
D_{\alpha}D_{\beta} \Omega - g_{\alpha \beta} D \cdot D \Omega
+ \frac{\gamma}{8} \left[g_{\alpha \beta} \frac{(D \Omega )^2 }
{\Omega} - 2 \frac{D_{\alpha } \Omega D_{\beta} \Omega}
{\Omega} \right] + \frac{\mu}{2}g_{\alpha \beta} \Omega ^{1-\lambda}
\label{eomg}
\end{equation}
$$
+ \frac{1}{8} (g_{\alpha \beta} F^2 -  4 g^{\mu \nu}
F_{\alpha \mu} F_{\beta \nu} )  \Omega^{1 - \epsilon /2}
- \frac{1}{4} g_{\alpha \beta}\Omega ^{\delta} (Df)^2
+ \frac{1}{2}\Omega ^{\delta} D_{\alpha } f D_{\beta} f
= 0,
$$
\begin{equation}
R + \frac{\gamma}{4} \left[ \frac{(D \Omega )^2}{\Omega^2} - 2
\frac{D \cdot D \Omega}{\Omega} \right] + (1-\lambda)\mu\Omega^{-\lambda}
+ \frac{1}{4} (1- \frac{\epsilon}{2})  \Omega^{- \epsilon /2} F^2 - 
\frac{\delta}{2} \Omega^{\delta -1} (Df)^2 = 0,
\label{eomo}
\end{equation}
\begin{equation}
\delta \Omega^{\delta-1}
D\Omega \cdot Df + \Omega^{\delta}  D\cdot Df =0,
\end{equation}
\begin{equation}
g^{\mu\alpha}g^{\nu\beta}\left[(1-\epsilon/2)D_{\nu}\Omega
 (D_{\alpha}A_{\beta}-D_{\beta}A_{\alpha})+\Omega(D_{\nu}D_{\alpha}
A_{\beta}-D_{\nu}D_{\beta} A_{\alpha}) \right] =0,
\end{equation}
where $\Omega=e^{-2\phi}$ and $D$ denotes the covariant derivative.
We choose to work in a conformal gauge as
$ g_{+-}=-e^{2\rho+\gamma\phi /2}/2$ and $g_{--}=g_{++}=0 $, partly to
simplify our analysis.  Moreover, when it comes to quantum aspects of
2-d dilaton gravity reported in literature, the conformal gauge choice has
been usual\cite{giddings}.  We also require the negative
signature for space-like coordinates and 
the positive signature for a time-like
coordinate.
Under this gauge choice, our original action, modulo total derivative terms,
can be written as 
\begin{equation}
I = \int dx^+ dx^- ( 4 \Omega \partial_+ \partial_- \rho
+ \frac{\mu}{2} e^{2 \rho} \Omega^{1-\lambda -\gamma/4}
+ \Omega^{\delta} \partial_+ f \partial_- f
 - e^{-2 \rho } \Omega^{(\gamma -2\epsilon)/4 +1} F^2_{-+} )
\label{conaction}
\end{equation}
where $F_{-+}=\partial_- A_+ - \partial_+ A_- $. The $\phi$ is included
deliberately
in the conformal factor to cancel the kinetic term for the dilaton field up to
a total derivative term, thereby helping the task of finding relevant
symmetries.

     The equations of motion under the conformal gauge are
\begin{equation}
\partial_+\partial_-\Omega + \frac{\mu}{4} e^{2 \rho} \Omega^{1-\lambda
-\gamma/4} +\frac{1}{2}e^{-2 \rho } \Omega^{(\gamma -2\epsilon)/4 +1} F^2_{-+}
=0,
\label{emrho}
\end{equation}
\begin{equation}
\partial_+\partial_-\rho + \frac{\mu}{8}(1-\lambda -\gamma/4)\frac{
e^{2 \rho}}{\Omega^{\lambda +\gamma/4}} + \frac{\delta}{4}\Omega^{\delta-1}
\partial_+ f \partial_- f
\label{emO}
\end{equation}
\[ - \frac{1}{4}((\gamma -2\epsilon)/4+1) e^{-2
\rho } \Omega^{(\gamma -2\epsilon)/4} F^2_{-+} =0,
\]
along with the equations for massless scalar field,
\begin{equation}
\delta(\partial_+\Omega\partial_-f + \partial_-\Omega\partial_+f)+2\Omega
\partial_+\partial_-f =0
\label{emf}
\end{equation}
and for gauge fields,
\begin{equation}
\partial_-(e^{-2 \rho } \Omega^{(\gamma -2\epsilon)/4 +1} F_{-+})=0,
\end{equation}
\begin{equation}
\partial_+(e^{-2 \rho } \Omega^{(\gamma -2\epsilon)/4 +1} F_{-+})=0.
\label{emgf}
\end{equation}
The equations for the abelian gauge field can be solved to give
\begin{equation}
F_{-+}=e^{+2 \rho } \Omega^{(-\gamma +2\epsilon)/4 -1} Q
\label{sgauge}
\end{equation}
where $Q$ is a constant.  In addition to the equations of motion, we
have to impose 
gauge constraints resulting from our choice of the conformal gauge. They are
given by 
\begin{equation}
\frac{\delta I}{\delta g^{\pm \pm}} = 0,
\end{equation}
where $I$ is the original action Eq.(\ref{oaction}). From the equations of motion
for the metric tensor Eq.(\ref{eomg}), we obtain the gauge constraints
\begin{equation}
\partial_{\pm}^2 \Omega - 2 \partial_{\pm} \rho \partial_{\pm} \Omega
+ \frac{1}{2} \Omega^{\delta} ( \partial_{\pm} f )^2 = 0.
\label{gconst0}
\end{equation}

     Now we have to find the static solutions in terms of equations of motion
Eq.(\ref{emrho})-(\ref{emgf}) with the gauge constraints Eq.(\ref{gconst0}).
The general static solutions can be found by assuming all functions except the
gauge field depend on a single space-like coordinate $x=x^+x^-$. Then from 
Eq.(\ref{sgauge}), we observe that the variable $F_{-+}$, originally
defined as  
$\partial_- A_+ - \partial_+ A_- $, automatically becomes dependent
only on $x$ and we can consistently reduce the partial differential equations
into the coupled second order ODEs. The
resulting ODEs are
\begin{equation}
x\ddot{\Omega}+\dot{\Omega} + \frac{\mu}{4} e^{2 \rho} \Omega^{1-\lambda
-\gamma/4} +\frac{1}{2}e^{-2 \rho } \Omega^{(\gamma -2\epsilon)/4 +1} F^2_{-+}
=0,
\label{eomrho}
\end{equation}
\begin{equation}
x\ddot{\rho}+\dot{\rho} + \frac{\mu}{8}(1-\lambda -\gamma/4)\frac{
e^{2 \rho}}{\Omega^{\lambda +\gamma/4}} + \frac{\delta}{4}\Omega^{\delta-1}
x\dot{f}^2 
\end{equation}
\[- \frac{1}{4}((\gamma -2\epsilon)/4+1) e^{-2 \rho }
\Omega^{(\gamma -2\epsilon)/4} F^2_{-+} =0,
\]
\begin{equation}
\Omega x \ddot{f}+\Omega\dot{f}+\delta \dot{\Omega}x\dot{f} =0,
\end{equation}
and
\begin{equation}
\frac{d}{dx}\left(e^{-2\rho} \Omega^{(\gamma -2\epsilon)/4 +1} F_{-+} \right)
=0,
\label{eoma}
\end{equation}
where the dot represents taking a derivative with respect to $x$.
The gauge constraints become
\begin{equation}
\ddot{\Omega}-2\dot{\rho}\dot{\Omega}+\frac{1}{2}\Omega^{\delta}\dot{f}^2=0.
\label{gconst}
\end{equation}
The general solutions of the above ODEs are the same as the general static
solutions of the original action under a particular choice of the conformal
coordinates. 
The above ODE's except the gauge constraint can also be derived
from an action
\begin{equation}
I = \int dx [ x \dot{\Omega} \dot{\rho}- \frac{\mu}{8}
e^{2 \rho}\Omega^{1-\lambda -\gamma/4}-\frac{1}{4} \Omega^{\delta} x \dot{f}^2
+\frac{1}{4} e^{-2 \rho} \Omega^{(\gamma -2\epsilon)/4 +1} \dot{A}^2
\label{xaction}
\end{equation}
$$- \frac{1}{2}\Omega^{3-\lambda -\epsilon /2} x(\dot{A} - F_{-+} )^2 ]
$$
by varying this action with respect to $\Omega$, $\rho$, $f$, $A$ and
$F_{-+}$. The field $A$ is introduced to get the ODE for $F_{-+}$, Eq.(\ref
{eoma}).  In fact, for $\rho$ and $\Omega$, we get the equations 
with $F_{-+}$ replaced by
$\dot{A}$ and additional terms containing $(\dot{A}-F_{-+})^2$. From the 
equation of motion for $F_{-+}$
\begin{equation}
\Omega^{3-\lambda -\epsilon /2}x(-\dot{A}+F_{-+})=0, 
\label{eomea}
\end{equation}
we find $\dot{A}=F_{-+}$ to eventually get the same 
equations for $\rho$ and $\Omega$
as before.
The ODE for $F_{-+}$, Eq.(\ref{eoma}), can be derived by the equation of
motion for $A$.

\subsection{Symmetries, Noether Charges and Explicit Solutions}

      We observe
that the action Eq.(\ref{xaction}) has four continuous symmetries
\begin{eqnarray*}
(a)&&f \rightarrow f + \alpha, \\
(b)&&A \rightarrow A + \alpha,\\
(c)&&x \rightarrow x e^{\alpha} ,~\rho \rightarrow \rho - \frac{1}{2}
\alpha ,~ F_{-+} \rightarrow F_{-+} e^{-\alpha}, \\
(d)&&x \rightarrow x^{1+\alpha} ,~  \rho \rightarrow \rho - \frac{1}{2}
(2-\lambda - \frac{\gamma}{4} ) \ln{(\alpha + 1)} - \frac{\alpha}{2} \ln{x},
~\Omega \rightarrow \Omega (1+\alpha ) ,\\
&&f \rightarrow f(1+\alpha )^ {(1-\delta )/2}, 
~A \rightarrow A (1+\alpha)^{(\epsilon+2\lambda-4)/4} ,
~F_{-+} \rightarrow F_{-+}x^{-\alpha} (1+\alpha)^{(2\lambda+\epsilon-8)/4}.
\end{eqnarray*}
Here $\alpha$ represents an arbitrary real parameter of each transformation.
The symmetries (a) and (b) are clear 
since field $f$ and $A$ appear only through
their derivative. The symmetry (c) is clearly a static remnant of the
underlying classical conformal invariance.  The only non-trivial symmetry
of our problem is (d).  This transformation, that changes the action by a
total derivative, is in fact a conformal coordinate transformation from 
$x^{\pm}$ to tortoise coordinates $\log x^{\pm}$ followed by some
overall scale transformation of relevant fields.
The Noether charges for these symmetries are straightforwardly constructed
as  
\begin{eqnarray*}
f_0&=&x\Omega^{\delta}\dot{f} \\
Q&=&e^{-2 \rho} \Omega^{(\gamma -2\epsilon)/4 +1} \dot{A}-2\Omega^{3-
\lambda -\epsilon /2} x(\dot{A} - F_{-+} ) \\
c_0&=&\frac{1}{2}x\dot{\Omega}+x^2\dot{\rho}\dot{\Omega}- \frac{1}{4}\Omega^
{\delta}x^2 \dot{f}^2 + \frac{\mu}{8}xe^{2\rho}\Omega^{1-\lambda-\gamma/4}
+ \frac{1}{4}e^{-2\rho} \Omega^{(\gamma-2\epsilon)/4+1}x \dot{A}^2 \\
& &- \frac{1}{2}\Omega^{3-\lambda -\epsilon /2} x^2(\dot{A}^2 - F^2_{-+} ) \\
s&=&-c_0\ln x-\frac{1}{2}x\dot{\Omega}(2-\lambda-\frac{\gamma}{4})
+x\dot{\rho}\Omega+\frac{\delta-1}{4}\Omega^{\delta}x\dot{f}f \\
& &+\frac{\epsilon+2\lambda-4}{8}e^{-2\rho}\Omega^{(\gamma-2\epsilon)/4+1}
\dot{A}A 
+ \frac{1}{2}\Omega - \frac{1}{4}(\epsilon+2\lambda-4)
\Omega^{3- \lambda -\epsilon /2}xA(\dot{A}-F_{-+}) ,
\end{eqnarray*}
respectively.
The imposition of the gauge constraint Eq.(\ref{gconst}) yields $c_0 =0$.
Furthermore, imposing
Eq.(\ref{eomea}), we simplify the Noether charges as
\begin{eqnarray}
f_0&=&x \Omega^{\delta} \dot{f} \label{f01} \\
Q&=&e^{-2\rho} \Omega^{(\gamma -2\epsilon)/4 +1} \dot{A} \label{a01} \\
0&=&x^2 \dot{\rho}\dot{\Omega}+\frac{1}{2}x\dot{\Omega} - \frac{1}{4}
\Omega^{\delta}x^2 \dot{f}^2 + \frac{\mu}{8}xe^{2\rho}\Omega^{1-\lambda-
\gamma/4} + \frac{1}{4}e^{-2\rho}\Omega^{(\gamma-2\epsilon)/4+1} x\dot{A}^2
\label{c01} \\
s&=&-\frac{1}{2}x\dot{\Omega}(2-\lambda-\frac{\gamma}{4}) +x\dot{\rho}\Omega
+\frac{\delta-1}{4}\Omega^{\delta}x\dot{f}f \nonumber \\
 & & +\frac{\epsilon+2\lambda-4}{8}
e^{-2\rho}\Omega^{(\gamma-2\epsilon)/4+1} \dot{A}A+ \frac{1}{2}\Omega 
\label{c11}
\end{eqnarray}
We can rewrite the equations of motion Eq.(\ref{eomrho})-(\ref{eoma})
in a form that represents the conservation of these Noether charges
$f_0$, $Q$, $c_0$ and $s$ ($c_0=0$ by the gauge constraint).
When we integrate Eqs.(\ref{f01})-(\ref{c11}) further to get closed form
solutions, we get four additional constants of integration. Among these 7
parameters, the meaning of $f_0$ and $Q$, the scalar charge and the $U(1)$
charge, respectively, are clear from the asymptotic behavior of the scalar
field and the $U(1)$ gauge field. Not all of the remaining 5 parameters are
physically important. We note that adding constant terms to $f$ and $A$ is
trivial. See, for example, Eq.(\ref{eomea}). Additionally, as we will
explain later, two parameters actually represent the degree of freedom in the
choice of coordinate systems, namely, the reference time choice and the
scale choice. These considerations show that the general static solutions
are parameterized by three parameters (including $Q$ and $f_0$), modulo
coordinate transformations.

We can explicitly demonstrate the structure of the solution space by
assuming $\delta=1$ and $2-\lambda-\gamma/2+\epsilon/2=0$. Moreover, in
our further consideration, we only consider the case when $Q>0$.\footnote{
If the $U(1)$ charge vanishes, the whole situation becomes identical to that
of Ref.\cite{kp}, where we already have a complete analysis. The results
for $Q<0$ can be trivially obtained from our results for $Q>0$.}
By this assumption, we exclude the case when $A$ field becomes degenerate,
being a strict constant.
By letting $\rho=\bar{\rho}+(2-\lambda-\gamma/4)(\ln{\Omega})/2-(\ln{x})/2$,
we get
\begin{equation}
(s - \frac{\epsilon+2\lambda-4}{8}Q A ) \dot{A}=Q
e^{2\bar{\rho}} \Omega^{2-\lambda-\gamma/2+\epsilon/2}\dot{\bar{\rho}}
\nonumber
\end{equation}
from Eq.(\ref{a01}) and Eq.(\ref{c11}). Here we see the role of the
assumption $2-\lambda-\gamma/2+\epsilon/2=0$.
For example, in the spherically symmetric reduction of $d$-dimensional
Einstein gravity, we have 
$\epsilon+2\lambda-4=-4(d-3)/(d-2)$ and $ 2-\lambda-\gamma/2+\epsilon/2=0$.
Under this condition, the above equation can be integrated directly to
yield
\begin{equation}
2s A +\frac{1}{2}(1+q)Q A^2 + c= Q e^{2\bar{\rho}}
\label{Aor}
\end{equation}
where $q+1=-(\epsilon+2\lambda-4)/4$ and $c$ is the constant of
integration.
From Eq.(\ref{f01}),(\ref{a01}) and (\ref{Aor}), we can determine $f$ via,
\begin{equation}
\dot{f}=\frac{f_0}{2s A +(1+q)Q A^2 /2 + c}\dot{A}
\label{fdot}
\end{equation}
which, upon integration, becomes
\begin{equation}
f=\frac{f_0}{\sqrt{4s^2-2(1+q)Q c }}\ln{\left| \frac{A-A_-}{A+A_+}
\right| } + f_1
\label{fs}
\end{equation}
where $A_\pm =(\sqrt{4s^2-2(1+q)Q c } \pm 2s)/[(1+q)Q]$ and $f_1$ is
the constant of integration. The constant of integration $f_1$ represents
the trivial constant term we can add to the scalar field $f$.
Using Eq.(\ref{f01}),({\ref{a01}) and (\ref{Aor}), we can rewrite
Eq.(\ref{c01}) as
\begin{equation}
0=4(1+q)\left(\frac{d\phi}{dA}\right)^2-2\frac{P'(A)}{P(A)}\frac{d\phi}{dA}-
\frac{f_0^2} {2P^2(A)}+\frac{2Q+\mu e^{-4(1+q)\phi}/Q}{4P(A)}
\label{OaA}
\end{equation}
where $P(A)= (1+q)QA^2/2+2sA+c$. The prime indicates the
differentiation with respect to $A$. By differentiating the above equation with
respect to $A$, we have
\begin{equation}
0=\left[ \frac{P^\prime}{P}-4(1+q)\frac{d\phi}{dA} \right]\left[ \frac{d^2\phi}
{dA^2}+2(1+q)\left(\frac{d\phi}{dA} \right)^2-\frac{f_0^2}{4P^2} \right].
\end{equation}
We see that two cases are possible.
When the first factor of the above equation is zero, we get,
\begin{equation}
|P(A)|=k \Omega^{-2(1+q)}
\end{equation}
where $k$ is the constant of integration which is greater than zero.
We must verify whether the above result is the true solution of Eq.(\ref{OaA}).
By substituting the above result into Eq.(\ref{OaA}) we get
\begin{equation}
0=D^2-\frac{(1+q) \mu}{Q}\frac{P}{|P|}k.
\end{equation}
where $D^2=4s^2-2(1+q)Qc+2(1+q)f_0^2$.
So we must fix the constant of integration as
\begin{equation}
k=\frac{D^2 Q}{\mu (1+q)} \frac{|P|}{P}>0
\end{equation}
We have found one solution
\begin{equation}
\Omega^{2(1+q)}=\frac{D^2 Q}{\mu (1+q)}\frac{1}{P}~~~.
\label{ops}
\end{equation}
In the second case where the second factor is zero,
we can find the solutions using the result derived in Appendix.
For $D^2 =4s^2-2(1+q)Q c+2(1+q)f_0^2 \neq 0$ we have
\begin{equation}
\phi=\frac{1}{4(1+q)} \left[ \ln{|P|}+2\ln{ \left| c_1e^{D I}-2(1+q)\right|}
-D I \right] + c_2
\label{ynz}
\end{equation}
where $c_1$ and $c_2$ are constants of integration and
$I(A)= \int P(A)^{-1}dA$.
Since the original equation Eq.(\ref{OaA}) is the first order equations we must
fix one of the constants in Eq.(\ref{ynz}). By plugging the above equation into
Eq.(\ref{OaA}) we get
\begin{equation}
e^{-4(1+q)c_2} = \frac{8 D^2 Q c_1}{-\mu}\frac{|P|}{P} > 0,
\label{s11}
\end{equation}
and thus, 
\begin{equation}
\Omega^{2(1+q)}=\frac{8D^2 Q c_1}{-\mu}\frac{e^{D I}}{P \left(c_1
e^{D I}-2(1+q) \right)^2}~~~.
\label{s1}
\end{equation}
For $D^2=0$ we have
\begin{equation}
\phi=\frac{1}{4(1+q)} \ln{ \left| P(I+c_1)^2 \right| } + c_2
\end{equation}
By substituting the above equation into Eq.(\ref{OaA}) we get
\begin{equation}
e^{-4(1+q)c_2}=-\frac{4Q}{\mu (1+q)} \frac{|P|}{P} > 0
\end{equation}
and
\begin{equation}
\Omega^{2(1+q)}=-\frac{4Q}{\mu (1+q)} \frac{1}{P(I+c_1)^2}.
\label{s2}
\end{equation}
Since $Q$ does not vanish, we can find $A$ as a function of $x$ by plugging
Eq.(\ref{Aor}) into Eq.(\ref{a01})
\begin{equation}
\ln{|x/x_0|}= \int \frac{\Omega(A)}{P(A)}dA
\label{Ax}
\end{equation}
where $x_0$ is the constant of integration. Thus, $x_0$ simply represents
the choice of the reference time. The following is a summary table
for solutions. 

\[
\begin{array}{c|c|c} \hline
 & D \neq 0 & D=0 \\
\hline
\rm{solutions} & \frac{D^2Q}{\mu(1+q)}\frac{1}{P} & \frac{-4Q}{\mu(1+q)}
 \frac{1}{P(I+c_1)^2} \\
\Omega^{2(1+q)} & \frac{8D^2Qc_1}{-\mu}\frac{e^{DI}}{P(c_1e^{DI}-2(1+q))^2}
& \\
\hline
\end{array}
\]
\[\rm{Table~for~Solutions}\]
\begin{eqnarray*}
D^2&=&4s^2-2(1+q)Qc+2(1+q)f_0^2 \\
P(A)&=&(1+q)QA^2/2+2sA+c \\
I(A)&=&\int P(A)^{-1}dA
\end{eqnarray*}

\subsection{Properties of Static Solutions}

The general solution space consists of some discrete number of 7
dimensional spaces. Among these parameters, the meaning of $f_0$ and $Q$
are clear since they are the scalar charge and the $U(1)$ charge. Additionally,
$x_0$ simply denotes the degree of freedom in choosing a reference time and
$f_1$ is a trivial addition of a constant term to the scalar field. This
leaves us with three parameter space $(c,c_1,s)$. To get the physical
degrees of freedom, we further notice that the transformation $A \rightarrow
A+k$ does not produce any physically distinguishable changes in the
solutions. The orbit of this transformation ($s \rightarrow s+(1+q)Qk/2$
and $c \rightarrow c -2sk -(1+q)Qk^2/2$ as can be seen from
Eq.(\ref{Aor})) should be modded out from the remaining three dimensional
space. Similarly, the scale transformation (corresponding to an arbitrary
choice of scale and related to the symmetry (c) in section 2.2) should also
be modded out. Thus, the physical solution space is parameterized by ($f_0$,
$Q$) and a parameter that parameterizes the coset of ($c$,$c_1$,$s$)
modulo two transformations above. This additional parameter corresponds to
the physical mass of a generic black hole. 

Given our general solutions, it is physically very interesting to see
what happens when a generic charged black hole tries to carry a scalar
hair. The Reissner-Nordstr\"{o}m type solutions are found in Eq.(\ref{s1}).
The range of $A$ is determined to be $A>A_-$ or $A<-A_+$, because $P(A)$,
which is equal to $(1+q)Q(A+A_+)(A-A_-)/2$, should be greater than zero
as can be seen in Eq.(\ref{Aor}). Since we have $\mu<0$ for most cases of
physical interests, we have $c_1 >0$ from
Eq.(\ref{s1}).
When the scalar charge $f_0$ vanishes, Eq.(\ref{s1}) becomes
\begin{equation}
\Omega^{2(1+q)}=\frac{2D^2 (k^2-1)}{-\mu (1+q)^2} \left[ A
+\frac{2s}{(1+q)Q}-\frac{D k}{(1+q)Q} \right]^{-2}
\end{equation}
where $k=[c_1+2(1+q)]/[c_1-2(1+q)]$, which satisfies $|k|>1$.
Since $\Omega$ is proportional to some power of 
the geometric radius of transversal
sphere, we require it to vary from 0 to infinity.
This gives us further restriction on the range of $A$ as
$A_-<A<(-2s+D k)/[(1+q)Q]$ if we assume $D >0$ and $k>1$.
The metric becomes
\begin{equation}
g_{+-}=-\frac{P(A)}{2Qx \Omega^{\epsilon/2}}
\end{equation}
where
\begin{equation}
P(A)=\frac{D^2 (k^2-1)}{2(1+q)Q} \left[ 1- \frac{2}{\Omega^{1+q}}
\frac{k}{\sqrt{k^2-1}}Q(-\mu/2)^{-1/2}
+ \frac{Q^2 (-\mu/2)}{\Omega^{2(1+q)}}\right].
\end{equation}
Note that $P(A_-)=0$, which shows $A_-$ is the outer horizon.
It is straightforward to verify 
that, for 4-dimensional spherically symmetric case,
our metric becomes that of the Reissner-Nordstr\"{o}m.

     If $f_0$ dose not vanish, 
the range of $A$ is determined
by Eq.(\ref{s1}) where
\begin{equation}
e^{D I}=\left|\frac{A-A_-}{A+A_+} \right|^{D/\sqrt{4s^2-2(1+q)Q
c}}.
\label{eI}
\end{equation}
Here we take the same assumption as $f_0=0$ case.
Since the exponent of Eq.(\ref{eI}) is greater than 1, we have
\[\lim_{A \rightarrow A_-^+} \Omega = 0, \]
which was a finite value in the Reissner-Nordstr\"{o}m case
($f_0 = 0$ case) representing the radius of the outer horizon.
The range of $A$ is $A_- < A < A_{\infty}$ where
\begin{equation}
A_{\infty}=\frac{1}{(1+q)Q} \left[ \sqrt{4s^2-2(1+q)Qc} 
\frac{1+\left(\frac{k-1}{k+1}\right)^{\sqrt{4s^2-2(1+q)Qc}/D}}
{1-\left(\frac{k-1}{k+1}\right)^{\sqrt{4s^2-2(1+q)Qc}/D}}
-2s \right]
\end{equation}
and \[\lim_{A \rightarrow A_{\infty}^-} \Omega = \infty. \]
The value $A_{\infty}$, that corresponds to the spatial infinity, becomes the
corresponding value of the Reissner-Nordstr\"{o}m type 
solutions if we set $f_0=0$.
Now let us see what happened to the outer horizon. From Eq.(\ref{s1}),
Eq.(\ref{a01}) and Eq.(\ref{Aor}), we get
\begin{equation}
2(1+q)\frac{d \Omega}{dy}=(1+q)Q \left[ \frac{D-2s}{(1+q)Q} -A \right]
+\frac{2D c_1 e^{D I}}{2(1+q)-c_1e^{D I}}
\label{dO2}
\end{equation}
where $y=\ln{x}$, an asymptotically flat coordinate near the
spatial infinity.
Since $e^{D I}$, Eq.(\ref{eI}), is a monotonically increasing function
of $A$ in $A_- <A<A_{\infty}$, we have
\begin{equation}
\lim_{A \rightarrow A_-^+} \frac{d \Omega}{dy}
= \left\{ \begin{array}{ll}
         0, & f_0=0 \\
         {\rm positive~value}, & f_0 \neq 0
        \end{array}
  \right.
\end{equation}
\begin{equation}
\lim_{A \rightarrow A_{\infty}^-} \frac{d \Omega}{dy}
= \infty.
\end{equation}
The result for $A \rightarrow A_-^+$ when $f_0=0$ is 
as expected since it just
shows $A_-$ corresponds to the outer (apparent) horizon.
When $f_0 \neq 0$, however, the apparent horizon does not  
exist between $A_-$ and
$A_{\infty}$, the physical range of $A$.  In fact, this
is non-trivial to verify, for 
if we define $F(A)$ as the right hand side of
Eq.(\ref{dO2}), we have $(dF/dA) |_{A=A_-} <0$.  Thus, it may
seem possible to
have $2(1+q) \frac{d \Omega}{dy}=F(A)=0$ between $A_-$ and
$A_{\infty}$. To investigate this possibility, we calculate the
minimum value of $F(A)$. At the minimum point $A=A_0$, we have
\begin{equation}
\frac{4 D(1+q)}{2(1+q)-c_1 e^{D I}} \Bigg\vert_{A=A_0}=
D + \sqrt{D^2+2(1+q)QP(A_0)} ,
\end{equation}
which follows from the condition $(dF/dA)|_{A=A_0}=0$.
The corresponding minimum value of $F(A)$ is given by
\begin{equation}
2(1+q)\frac{d \Omega}{dy}\Bigg\vert_{A=A_0}=
\sqrt{2(1+q)f_0^2 + [2s+(1+q)QA_0]^2}-2s-(1+q)QA_0
\end{equation}
which is greater than zero if $f_0 \neq 0$.
Therefore, we have $d\Omega /dy >0$ in the specified 
physical range of $A$, which in turn implies there is no
apparent horizon in the same range of $A$.
Note that at $A=A_-$, 
where $\Omega=0$, the scalar field, Eq.(\ref{fs}), diverges
logarithmically. This shows that the would-be horizon is 
shielded by a naked
singularity produced by the diverging scalar field (if $f_0 \ne 0$),
just like the case of electrically neutral black holes \cite{kp}.
This consideration holds in all (not 
just 4-d Reissner-Nordstr\"{o}m) model
theories we consider and, thus, illustrates
``no-scalar-hair property'' \cite{chase}.

It is clear that our solutions are defined on a local coordinate patch. In
the process of getting general static solutions, we found many solutions
(branches) that show markedly different behavior from the space-time
geometry with asymptotically flat Minkowskian geometry. It will be an
interesting exercise to glue them together to construct non-trivial and
physically interesting global space-time structures.

\begin{flushleft}{\Large \bf Acknowledgments}
\end{flushleft}

     The earlier version of this work was completed while Y.K. was
at Physics Department of Princeton University.  Y.K. wishes to thank
H. Verlinde for useful discussions. D. Park wishes to thank H.C. Kim
for useful discussions.

\appendix
\begin{flushleft}{\Large \bf Appendix}
\end{flushleft}

In this section we will find the general solutions of the following second
order nonlinear ordinary equations
\begin{equation}
\frac{d^2 y}{dx^2}-h \left(\frac{dy}{dx}\right) ^2=\frac{g}{(ax^2
+bx+c)^2}
\label{snODE}
\end{equation}
where $a$,$b$,$c$ and $g$ are constants, and $h$ is a nonzero constant.
Let $z(x)=y ^\prime \equiv dy/dx $ and $P(x)=ax^2 +bx+c$. Since $P(x)$ is a
second order polynomial, we can guess one simple solution of $z(x)$ as
$z_0 (x)= (c_1 x+c_2)/P(x)$. In fact, if we let $z_0(x)=-(P^\prime + D)/
(2h P)$, we find
\begin{equation}
z_0 ^\prime - h z_0 ^2 = \frac{b^2 -4ac-D^2}{4h}
\frac{1}{P^2(x)}.
\end{equation}
The equality $D^2 =b^2-4ac-4 h g$ confirms that $z_0(x)$ is a 
possible solution.
To find the general solutions, let $z(x)=z_0(x)+v(x)$.  
Then we get the equation
for $v(x)$
\begin{equation}
v^\prime -h v^2 -2 h z_0(x) v =0.
\end{equation}
By letting $u(x)=P(x)v(x)$, we get
\begin{equation}
\frac{u^\prime}{h u^2 -D u} = \frac{1}{P(x)} ,
\end{equation}
which can be integrated easily.

\vspace{1cm}
\pagebreak[3]
\addtocounter{subsection}{1}
\noindent{ \large \bf $D \neq 0$ case}
\nopagebreak
\vspace{2mm}
\nopagebreak

In this case, we get 
\begin{equation}
z(x)=-\frac{1}{2h} \left\{ \frac{P^\prime(x)}{P(x)}-\frac{D \left[
h \mp  c_1 e^{D I(x)} \right]}{P(x) \left[ h\pm c_1
e^{D I(x)}
\right]} \right\}
\end{equation}
for $z(x)$ 
where $I(x)=\int P(x)^{-1} dx$ and $c_1$ is a non-negative constant. After
the integration over $x$, we finally get
\begin{equation}
y(x)=-\frac{1}{2h} \left[ \ln{\left| P(x) \right| }+2\ln{\left| h
\pm c_1 e^{D I(x)} \right| }- D I(x) \right] + c_2.
\end{equation}
where $c_2$ is a constant of integration. The minus sign in front of $c_1$
can be absorbed into $c_1$ so that $c_1$ can be made less than zero.

\vspace{1cm}
\pagebreak[3]
\addtocounter{subsection}{1}
\noindent{ \large \bf $D = 0$ case}
\nopagebreak
\vspace{2mm}
\nopagebreak

In this case, we simply get
\begin{equation}
y(x)=-\frac{1}{2h} \ln{ \left| P(x) \left[I(x)+c_1 \right]^2 \right|}
+ c_2
\end{equation}
where $c_1$and $c_2$ are constants.
Using an elementary method, $I(x)$ is calculated to be
\begin{equation}
I(x)=\int \frac{dx}{ax^2+bx+c} = \frac{1}{\sqrt{b^2-4ac}} \ln{ \left|
\frac{\sqrt{b^2-4ac}-b-2ax}{\sqrt{b^2-4ac}+b+2ax} \right| }
\end{equation}
for $b^2-4ac > 0$.

\end{document}